\documentclass[referee]{raa_rb}   
\usepackage{graphicx,times}
\usepackage{natbib}
\usepackage{amssymb,amsmath}
\usepackage{adjustbox}
\usepackage{rotating}
\usepackage{lscape}
\usepackage{pdflscape}
\usepackage{verbatim}
\usepackage{bm}
\bibpunct{(}{)}{;}{a}{}{,}
\usepackage[a4paper=true,dvipdfm=true,pagebackref=true]{hyperref}
\usepackage[a4paper, total={6in, 9.5in}]{geometry}
\hypersetup{pdftitle = The title of my PDF, pdfauthor = My name, pdfsubject= The subject, pdfkeywords = keyword1 keyword2 keyword3} 
\hypersetup{colorlinks = true, linkcolor = green, anchorcolor = red, citecolor = blue, filecolor = red, pagecolor = red, urlcolor = red}

\begin{document}

   \title{An Improved Pair Method to Probe the Dust Extinction Law
}

 \volnopage{ {\bf 20XX} Vol.\ {\bf X} No. {\bf XX}, 000--000}
   \setcounter{page}{1}

   \author{Yuxi Wang\inst{1,2,3}, Jian Gao\inst{1,3}, Yi Ren\inst{2,3}, Jun Li\inst{3}
   }

   \institute{ Institute for Frontiers in Astronomy and Astrophysics, Beijing Normal University, Beijing 102206, China; {\it jiangao@bnu.edu.cn}\\
   \and 
   College of Physics and Electronic Engineering, Qilu Normal University, Jinan 250200,China; {\it yiren@qlnu.edu.cn}\\
        \and
        Department of Astronomy, Beijing Normal University, Beijing 100875, 
        China;\\
\vs \no
   {\small Received 20XX Month Day; accepted 20XX Month Day}
}

\abstract{Dust extinction law is crucial to recover the intrinsic energy distribution of celestial objects and infer the characteristics of interstellar dust.
Based on the traditional pair method, an improved pair method is proposed to model the dust extinguished spectral energy distribution (SED) of an individual star.
Instead of the mathematically parameterizing extinction curves, the extinction curves in this work are directly from the silicate-graphite dust model, so that the dust extinction law can be obtained and the dust properties can be analyzed simultaneously.
The ATLAS9 stellar model atmosphere is adopted for the intrinsic SEDs in this work, while the silicate-graphite dust model with a dust size distribution of $dn/da \sim a^{-\alpha}{\rm exp}(-a/a_c),~0.005 < a < 5~\mu{\rm m}$ for each component is adopted for the model extinction curves.
One typical extinction tracer in the dense region (V410 Anon9) and one in the diffuse region (Cyg OB2 \#12) of the MW are chosen to test the reliability and the practicability of the improved pair method in different stellar environments. 
The results are consistent with their interstellar environments and are in agreement with the previous observations and studies, which prove that the improved pair method is effective and applicable in different stellar environments.
In addition to the reliable extinction results, the derived parameters in the dust model can be used to analyze the dust properties, which cannot be achieved by other methods with the mathematical extinction models.
With the improved pair method, the stellar parameters can also be inferred and the extinction law beyond the wavelengths of observed data can be predicted based on the dust model as well.
\keywords{ISM: dust, extinction ---
stars: individual (V410 Anon9, Cgy OB2 \#12)
}
}

   \authorrunning{Wang et al.}            
   \titlerunning{An Improved Pair Method to Probe the Dust Extinction Law}  
   \maketitle

%
\section{Introduction}           
\label{sec:introduction}
The wavelength dependence of interstellar extinction, which is known as the extinction law or extinction curve,  is one of the primary sources of information about the interstellar grain population \citep{2003ARA&A..41..241D}. 
Extinction curves from the ultraviolet (UV) to near infrared (NIR) bands in the Milky Way (MW) can be characterized by the total-to-selective extinction ratio $R_V$ [$R_V \equiv A_V/E(B-V)$] (\citealt{1989ApJ...345..245C}, hereafter CCM).
The value of $R_V$ depends on the interstellar environment along the sight line. 
The large value of $R_V$ corresponds to the region with high density.
For instance, denser clouds tend to have larger values of $R_V$ ($\approx 5.5$) and the typical diffuse interstellar medium usually has a smaller one ($R_V \approx 3.1$). 

The mathematical model is a widely-used way to determine the extinction law.
In addition to the CCM model, the \citet[hereafter F90]{1990ApJS...72..163F} model contains 6 parameters and can be used to model the shape of extinction curves in UV bands (see eq.2 in FM90).
\citet[hereafter F99]{1999PASP..111...63F} estimated new $R_V$-dependent extinction curves in the MW from UV to IR  based on the relationship between $R_V$ and the parameters in the FM90 model, and the F99 model has been updated as the \citet[hereafter F04]{2004ASPC..309...33F} model.

Although the parameterized extinction models are effective ways to derive the extinction law, they are derived by mathematical parameter fitting rather than based on the physical model.
In addition, the mathematical extinction models are not universal in all galaxies, and one model can be only applied to one or some specific galaxies.
For example, 
the MW-type extinction law does not generally apply to the interstellar dust in other galaxies \citep{2015ApJ...815...14C}, and the single parameter $R_V$ cannot be used to completely describe the extinction feature in the extinction curves in external galaxies \citep{2022ApJS..259...12W}.
Instead, the dust model with the physical basis is another method to obtain the extinction curves. 
As an extinction curve derived from the dust model presented, the properties of dust can also be obtained, which can not be analyzed directly from the parameterized extinction curves.

The traditional ``pair method'' \citep{1970IAUS...36...28B}, which compares the spectrum of a reddened star with that of an unreddened one with the same or similar spectral type, is universally adopted to determine the dust extinction law in star-resolved galaxies.
In this work, the silicate-graphite dust model is adopted to propose an improved pair method, so that the dust extinction law can be obtained and dust properties can be analyzed simultaneously. 
One extinction tracer located in the dense region of the MW and one located in the diffuse region of the MW are adopted to test the practicability and the reliability of the improved method.
The details of the method are described in Section \ref{sec:method}, and the extinction tracers are introduced in Section \ref{sec:tracers}.
Section \ref{sec:re} shows our results and discussions.
We finally summarize our method in Section \ref{sec:summary}.

\section{Method}
\label{sec:method}
Based on the traditional pair method mentioned above, we propose an improved pair method that models the spectral energy distributions (SEDs) in combination with the intrinsic SEDs obtained from the stellar model atmosphere extinguished by the model extinction curves derived from the dust model.
With the model SEDs fitting to the observed data, the dust extinction law can be derived, the dust properties can be further analyzed based on the derived parameters in the dust model, and even the stellar parameters can be inferred from the stellar model atmosphere.

\begin{figure} 
  \centering
  \includegraphics[width=15cm, angle=0]{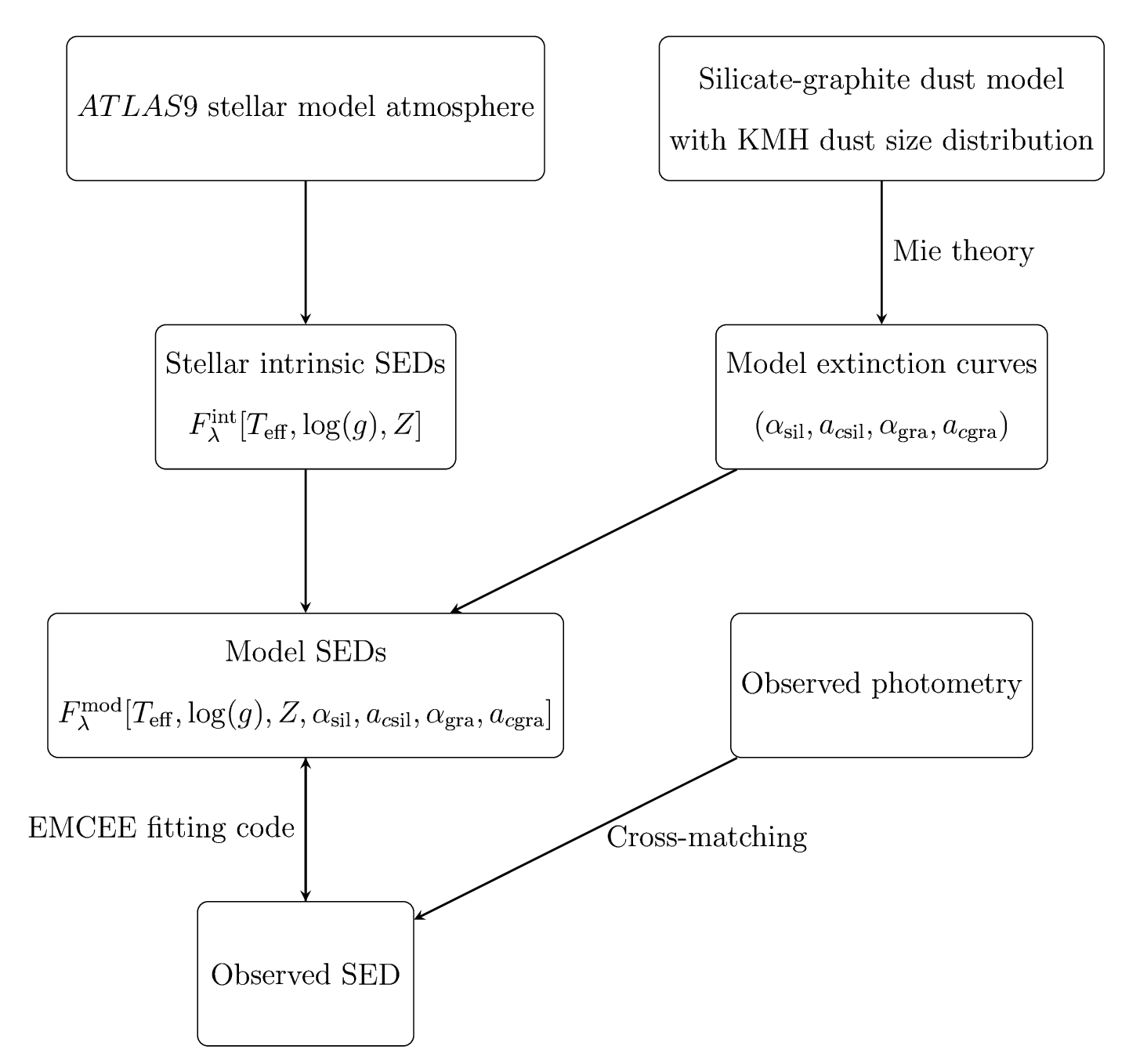}
  \caption{The improved pair method.
  The intrinsic SEDs from the ATLAS9 stellar model atmosphere and the model extinction curves from the dust model are introduced in Section \ref{subsec:int_sed} and Section \ref{subsec:ext}, respectively.
  The construction of model SEDs and the method that fits the model SEDs to the observed data are presented in Section \ref{subsec:fit}.
} 
  \label{fig:method}
  \end{figure}

Figure \ref{fig:method} summarizes the improved pair method proposed in this work, and the following subsections show the details.

\subsection{The intrinsic SEDs from the ATLAS9 stellar model atmosphere}\label{subsec:int_sed}

The traditional pair method considers the spectrum of an unreddened star with the same spectral type as the intrinsic spectrum.
However, sometimes it is not easy to obtain the intrinsic spectra from real stars especially in external galaxies because of the limited sample of unreddened standard stars.
The stellar model atmosphere can also be used to obtain the standard spectrum, which is at least as accurate as those from the real stars \citep{2005AJ....130.1127F}.

We adopt the ATLAS9 stellar model atmosphere to obtain the intrinsic SEDs in this work.
The code is written by \citet{1970SAOSR.309.....K} and is used for computing the local thermodynamic equilibrium blanketed stellar atmosphere model.
The ATLAS9 grid considers effective temperature $T_{\rm eff}$ from 3500 K to 50000K, surface gravity log$(g)$ from 0.0 to 5.0 and several chemical compositions.


\subsection{The model extinction curves from the dust model} \label{subsec:ext}

As mentioned in Section \ref{sec:introduction}, the extinction curves derived from the dust model are applied in this work.
Along the sight line, the extinction at the wavelength of $\lambda$ can be expressed as:
\begin{equation} \label{eq:3}
    A_{\lambda}/N_{\rm H} = 1.086 \int_{a_{\rm min}}^{a_{\rm max}} C_{\rm ext}(a, \lambda) \frac{1}{n_{\rm H}} \frac{dn}{da} da,
\end{equation}
where $a$ is the radius of the dust grain, $C_{\rm ext}(a,\lambda)$ is the extinction section and can be calculated by Mie theory \citep{1984ApJ...285...89D}, $n_{\rm H}$ is the number density of H nuclei, $N_{\rm H}$ is the column density of H nuclei, and $dn/da$ is the dust size distribution.

In terms of dust composition, there are mainly the silicate-graphite [+ polycyclic aromatic hydrocarbons (PAH)] model, the COMP model (silicate, graphite, PAH, amorphous carbon particles, organic refractory material, water ice and voids, \citealt{2004ApJS..152..211Z}), the silicate core-carbonaceous mantle model \citep{1978codu.book..187G,1997A&A...323..566L} and the composite dust model \citep{1989ApJ...341..808M}.
Three kinds of dust distributions are widely adopted for the silicate-graphite dust model: the power-law size distribution ($dn/da \sim a^{-\alpha}$) proposed by Mathis, Rumpl, and Nordsieck (hereafter MRN, \citealt{1977ApJ...217..425M}), the exponentially cutoff power-law size distribution [$dn/da \sim a^{-\alpha}{\rm exp}(a/a_c)$] proposed by Kim, Martin and Hendry (hereafter KMH, \citealt{1994ApJ...422..164K}), and the two log-normal size distributions for two population of PAHs \citep{2001ApJ...554..778L} adding to the traditional silicate-graphite dust model \citep{2001ApJ...548..296W}.

In this work, we adopt the typical silicate-graphite dust model\footnote{According to \citet{2011piim.book.....D}, a class of models that has met with some success assumes the dust to consist of two materials: amorphous silicate and carbonaceous materials. 
\citet{1977ApJ...217..425M} showed that models using silicate and graphite spheres could reproduce the observed extinction from the near-IR to the UV. 
\citet{1984ApJ...285...89D} presented self-consistent dielectric functions for graphite and silicate and showed that the silicate-graphite model appeared to be consistent with what was known about dust opacities in the far-IR. 
In addition, the silicate-graphite dust model has been widely used in modeling the dust extinction laws (e.g., MW: \citealt{1994ApJ...422..164K}, \citealt{2015ApJ...811...38W}; Magellanic Clouds: \citealt{2003ApJ...594..279G}; SN 2014J: \citealt{2015ApJ...807L..26G}; SN 2012cu: \citealt{2020P&SS..18304627G}; M31: \citealt{2022ApJS..259...12W}; M33: \citealt{2022ApJS..260...41W}; SN 2010jl: \citealt{2022MNRAS.511.2021L}) and is the basis of some famous works related to interstellar dust (e.g., \citealt{2001ApJ...548..296W}; \citealt{2004ApJS..152..211Z}, etc). } that consists of spherical amorphous silicate and graphite \citep{1984ApJ...285...89D} and choose the KMH dust size distribution for both components
: $dn/da \sim a^{-\alpha}{\rm exp}(a/a_c)$ in the size range of $0.005 < a < 5~{\mu}{\rm m}$, where a is the spherical radius of the dust, ${\alpha}$ is the power index, and $a_c$ is the exponential cutoff size.


\subsection{Compare model SEDs with the observed data}\label{subsec:fit}

The observed SED $F^{\rm obs}_{\lambda}$ of a reddened star can be expressed as \citep{2005AJ....130.1127F}:
\begin{equation}
F^{\rm obs}_{\lambda}=F^{\rm int}_{\lambda}{\theta}_{R}^2 10^{-0.4A_{\lambda}},
\end{equation}
where $F^{\rm int}_{\lambda}$ is the intrinsic surface flux of the star at wavelength ${\lambda}$, ${\theta}_R \equiv (R/d)^2$ is the angular radius of the star (where $d$ is the distance and $R$ is the stellar radius), and $A_{\lambda}$ is the absolute attenuation of the stellar flux by intervening dust at $\lambda$.
With the combination of the intrinsic SEDs and the model extinction curves obtained in the previous two subsections, a grid of model SEDs [$F^{\rm mod}_{\lambda}(T_{\rm eff}, {\rm log}~g, Z, {\alpha}_{\rm sil}, a_{c\rm sil}, {\alpha}_{\rm gra}, a_{c\rm gra})$] can be constructed, where $T_{\rm eff}$, log $g$, and $Z$ are, respectively, the effective temperature, the surface gravity, and the metallicity of the extinction tracer, while $\alpha_{\rm sil}$ ($\alpha_{\rm gra}$) and $a_{c\rm sil}$ ($a_{c \rm gra}$) are the power index and the exponential cutoff size in the dust size distribution [$dn/da \sim a^{-\alpha}{\rm exp}(a/a_c)$] of silicate (graphite).

Since the model SEDs for the tracers are pre-constructed before the calculation, it is only necessary to compare the observed data with the model SEDs in the template library to obtain the dust extinction law.
In the application of the method proposed in this work, some prior knowledge (e.g., the stellar parameters, the spectral type of the extinction tracer, etc) will constrain the range of the parameters and reduce the number of the model SEDs used for comparison, which simplifies the calculation, while the composition of the parameters can be adjusted to balance the complexity of the model and the quality of the fitting. 

The EMCEE fitting code \citep{2013PASP..125..306F} is adopted in this work to fit the model SEDs to the observed data.
It is a Markov Chain Monte Carlo (MCMC, \citealt{2010CAMCS...5...65G}) ensemble sampler based on the Bayesian probability theory \citep{1946AmJPh..14....1C} and helps with the most suitable parameters and the corresponding confidence intervals.
We apply the Gaussian likelihood and impose a flat prior on the fitting parameters for the extinction tracers.

We also use the Levenberg-Marquardt method \citep{2001ApJ...548..296W,2015MNRAS.453.3300A,2016P&SS..133...36N} for a comparison:

\begin{equation}
\chi^2/d.o.f.=\frac{1}{N_{\rm data}-N_{\rm para}}\sum \frac{({\rm log}(f_{\rm model})-{\rm log}(f_{\rm observed}))^2}{\sigma^2},
\end{equation}
where $N_{\rm data}$ is the number of the observed photometric points, $N_{\rm para}$ is the number of adjustable parameters, $f_{\rm observed}$ is the observed flux of each photometric point, $f_{\rm model}$ is the model flux of each photometric point, and $\sigma$ is the difference between the logarithm of extreme and logarithm of $f_{\rm observed}$. 
The model SED that fits the observed data best can be found by minimizing the $\chi^2/d.o.f.$, and the corresponding model extinction curve can be regarded as the extinction law and the dust property can be obtained simultaneously.


\section{The extinction tracers}
\label{sec:tracers}

\subsection{The extinction tracer in the dense region}

The main-sequence star named V410 Anon9 is chosen as the extinction tracer in the dense region of the MW.
It is located in the vicinity of an actively star-forming cloud (L1495E) within the Taurus-Auriga cloud complex \citep{1994ApJ...424..237S}. 
Its spectral type is A2 and the extinction in the V band is $A_V \approx 10$ mag \citep{1994ApJ...424..237S,2008ApJS..176..457S}, while other members in this cloud are nearly all later than M3 spectral type and their magnitudes of extinction in the V band are much lower. 
The relatively high brightness and reddening of V410 Anon9 make it a typical extinction tracer for the dense region.


\subsubsection{The observed SED}\label{subsubsec:v410_obs}

\begin{figure}
      \centering
     \includegraphics[width=14.5cm, angle=0]{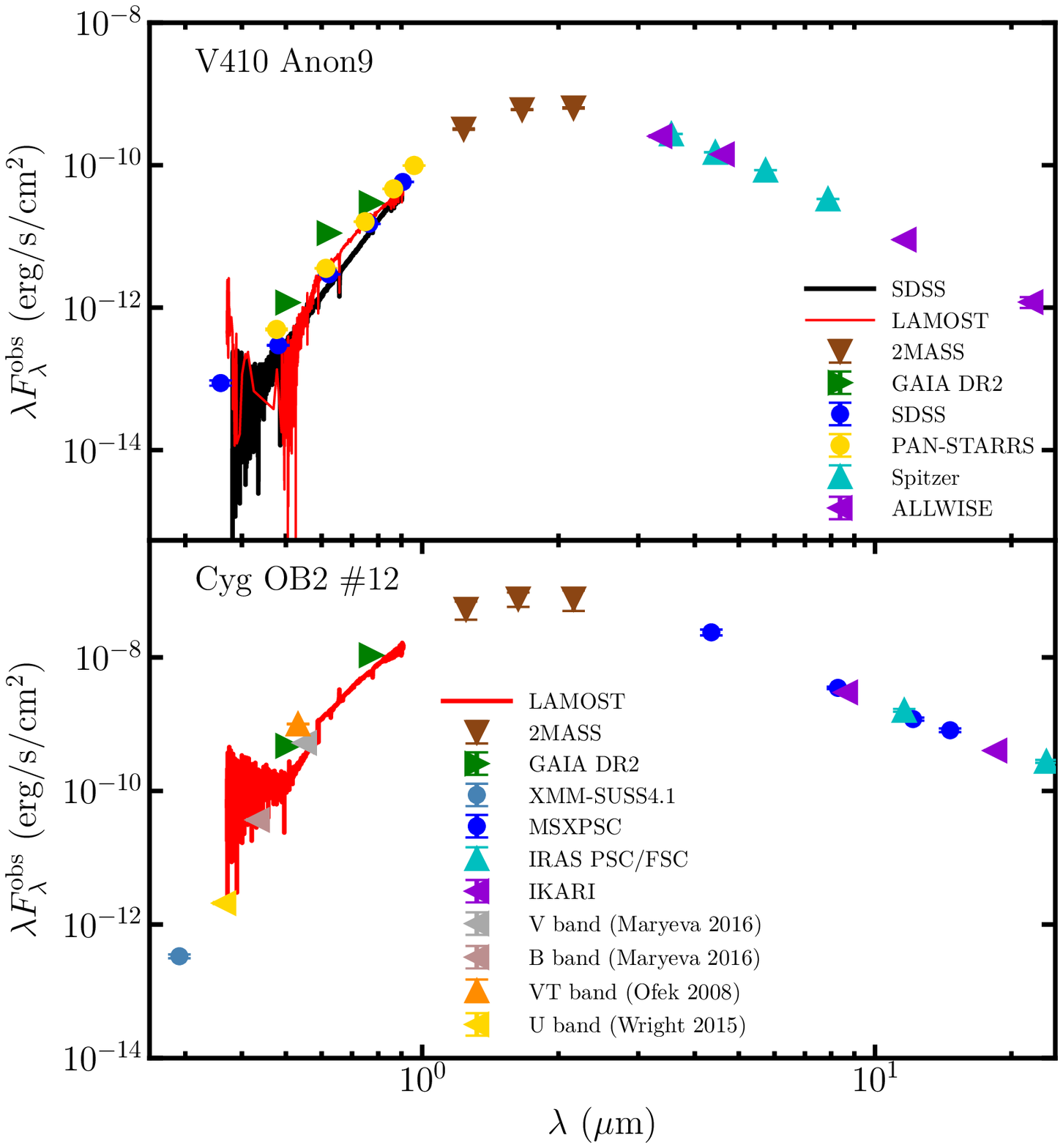}
      \caption{The observed SEDs of V410 Anon9 and Cyg OB2 \#12. We only adopt the observed photometry to derive the extinction laws. \label{fig:sed}}
      \end{figure}

In this work, we combine the spectroscopic data and the photometric data to construct the observed SEDs of V410 Anon9, which is shown in the upper panel of Figure \ref{fig:sed}.
The spectra are from the Sloan Digital Sky Survey (SDSS, \citealt{2019ApJS..240...23A}) and the Large Sky Area Multi-Object Fiber Spectroscopic Telescope (LAMOST, \citealt{2012RAA....12..723Z, 2012RAA....12.1197C}).
The photometric data, which are listed in Table \ref{tab:photometry}, are from Panoramic Survey Telescope And Rapid Response System (Pan-STARRS, \citealt{2016arXiv161205560C}) in $g, r, i, z, y$ bands, the Two Micron All Sky Survey (2MASS, \citealt{2003yCat.2246....0C}) in $J, H, K_S$ bands, the Spitzer/Galactic Legacy Infrared Midplane Survey Extraordinaire (Spitzer/GLIMPSE, \citealt{2009PASP..121..213C}) in $[3.6], [4.5], [5.8], [8.0]$ bands, the Wide-field Infrared Survey Explore (WISE, \citealt{2013wise.rept....1C}) in $W_1, W_2, W_3. W_4$ bands and GAIA DR2 \citep{2018yCat.1345....0G} in $G_{BP}, G, G_{RP}$ bands.

However, the original spectra from the LAMOST are uncalibrated.
In addition, the extinction curves in the bands covered by the LAMOST spectra and the SDSS spectra (optical bands) do not show the special features as the absorptions in the UV and the IR bands, and the observations in the optical bands cannot restrict the shape of the extinction curves and the dust properties as strongly as those in the UV and IR bands.
As a result,
only photometric data are adopted to calculate the extinction law.
It should be noted that because of the lack of uncertainties in Spitzer/GLIMPSE bands and the quite wide wavelength range of Gaia DR2 $G$ band\footnote{There will be large uncertainty in determing the effective wavelength of the $G$ band because of the wide wavelength range.}, we cannot apply these photometric data to determine the extinction curves.

  \begin{table}
    \bc
    \caption[]{The observed photometry of V410 Anon9 and Cyg OB2 \#12.\label{tab:photometry}}
    \setlength{\tabcolsep}{3pt}
     \begin{tabular}{ccccccccccccc}
      \hline\noalign{\smallskip}
      &  & V410 Anon9 &  & Cyg OB2 \#12 &  & \\
      Band & $\lambda_{\rm eff}$ & Magnitude & Uncertainty & Magnitude & Uncertainty & Ref$^a$ \\
       & $(\mu {\rm m})$ & (mag) & (mag) & (mag) & (mag) & \\
      \hline\noalign{\smallskip}
      $u$ & 0.359 & 21.389 & 0.093 & - & - & [1] \\
      $g$ & 0.481 & 19.710 & 0.014 & - & - & [1] \\
      $r$ & 0.623 & 16.937 & 0.004 & - & - & [1] \\
      $i$ & 0.764 & 14.945 & 0.004 & - & - & [1] \\
      $z$ & 0.906 & 13.269 & 0.004 & - & - & [1] \\
      $g$ & 0.477 & 19.161 & 0.028 & - & - & [2] \\
      $r$ & 0.613 & 16.744 & 0.007 & - & - & [2] \\
      $i$ & 0.748 & 14.893 & 0.002 & - & - & [2] \\
      $z$ & 0.865 & 13.584 & 0.006 & - & - & [2] \\
      $y$ & 0.960 & 12.648 & 0.010 & - & - & [2] \\
      $J$ & 1.235 & 10.205 & 0.026 & 4.667 & 0.324 & [3] \\
      $H$ & 1.630 & 8.717 & 0.020 & 3.512 & 0.260 & [3] \\
      $K_S$ & 2.160 & 7.908 & 0.024 & 2.704 & 0.364 & [3] \\
      $[3.6]$ & 3.550 & 7.330 & - & - & - & [4] \\
      $[4.5]$ & 4.439 & 7.260 & - & - & - & [4] \\
      $[5.8]$ & 5.731 & 7.140 & - & - & - & [4] \\
      $[8.0]$ & 7.872 & 7.140 & - & - & - & [4] \\
      $W1$ & 3.350 & 7.580 & 0.031 & - & - & [5] \\
      $W2$ & 4.600 & 7.235 & 0.019 & - & - & [5] \\
      $W3$ & 11.600 & 7.303 & 0.020 & - & - & [5] \\
      $W4$ & 22.100 & 7.435 & 0.186 & - & - & [5] \\
      $G$ & 0.623 & 15.384 & 0.001 & 8.889 & 0.003 & [6] \\
      $G_{BP}$ & 0.505 & 18.127 & 0.010 & 11.619 & 0.006 & [6] \\
      $G_{RP}$ & 0.772 & 13.850 & 0.005 & 7.441 & 0.005 & [6] \\
      $UVW1$  & 0.291 & - & - & 20.121 & 0.073 & [7] \\
      $U$ & 0.365 & - & - & 17.180 & 0.034 & [8] \\
      $B$ & 0.433 & - & - & 14.760 & 0.026 & [9] \\
      $V$ & 0.550 & - & - & 11.500 & 0.017 & [9] \\
      $D$ & 14.650 & - & - & 1.660 & 0.069 & [10] \\
      $C$ & 12.130 & - & - & 1.847 & 0.071 & [10] \\
      $A$ & 8.280 & - & - & 1.905 & 0.045 & [10] \\
      $B2$ & 4.350 & - & - & 1.885 & 0.155 & [10] \\
      $IRAS 12$ & 11.600 & - & - & 2.101 & 0.037 & [11] \\
      $IRAS 25$ & 12.900 & - & - & 1.649 & 0.055 & [11] \\
      $S9W$ & 8.610 & - & - & 1.942 & 0.018 & [12] \\
      $L18W$ & 18.400 & - & - & 1.678 & 0.009 & [12] \\
      $VT$ & 0.532 & - & - & 18.810 & 0.010 & [13] \\
      \noalign{\smallskip}\hline
    \end{tabular}
    \ec
    \tablecomments{0.65\textwidth}{$^a$ [1]: \citet{2019ApJS..240...23A}, [2]: \citet{2016arXiv161205560C}, [3]: \citet{2003yCat.2246....0C}, [4]: \citet{2009PASP..121..213C}, [5]: \citet{2013wise.rept....1C}, [6]: \citet{2018yCat.1345....0G}, [7]: \citet{2019yCat.2356....0P}, [8]: \citet{2015MNRAS.449..741W}, [9]: \citet{2016MNRAS.458..491M}, [10]: \citet{2003yCat.5114....0E}, [11]: \citet{2015AandC....10...99A}, [12]: \citet{2010AandA...514A...1I}, [13]: \citet{2008PASP..120.1128O}.}
    \end{table}

\subsubsection{The model SEDs}\label{subsubsec:v410_mod}

\begin{figure}
  \centering
 \includegraphics[width=14.5cm, angle=0]{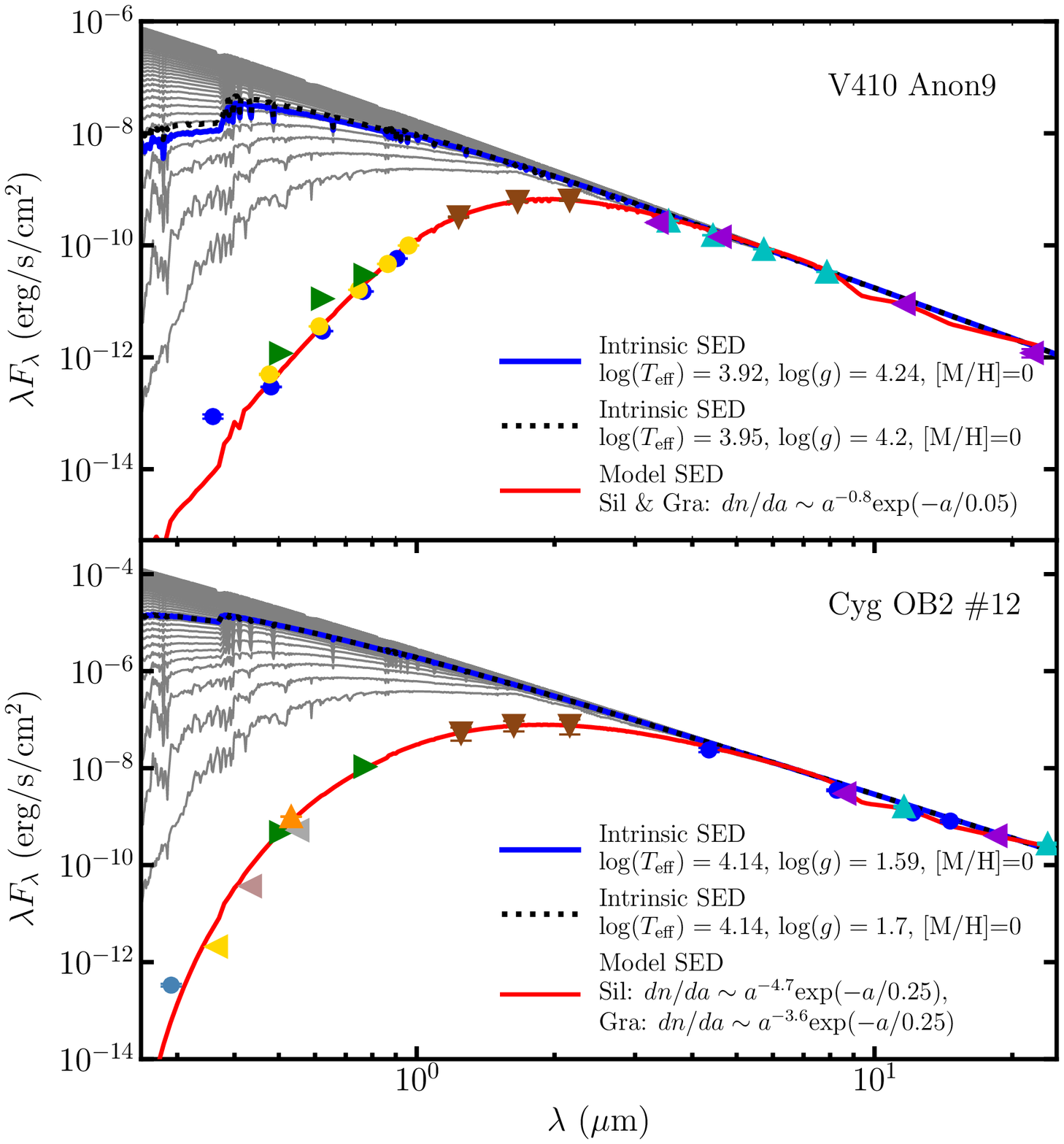}
	\caption{The intrinsic SEDs (gray, black and blue lines) derived from the ATLAS9 stellar model atmosphere for V410 Anon9 and Cyg OB2 \#12 as well as the best-fitting SEDs compared with the observed photometry (red solid line).
  The gray solid lines in the upper panel are the intrinsic SEDs from the ATLAS9 stellar model atmosphere with different effective temperature values and surface gravity values, while the blue solid line is the intrinsic SED with the derived stellar parameters (see Section \ref{subsec:st_ext} for details).\label{fig:F_f}}
  \end{figure}

For the main sequence tracer V410 Anon9, we construct the model intrinsic SEDs with 47 values of the effective temperature ($4000 \le T_{\rm eff} \le 50000$ K with 1000 K steps), 17 values of surface gravity ($1.00 \le {\rm log}(g) \le 5.00$ dex with 0.25 dex steps) and the solar abundances.
The intrinsic SEDs from the ATLAS9 stellar model atmosphere are plotted in the upper panel of Figure \ref{fig:F_f} with gray solid lines.
The black dotted line in the upper panel of Figure \ref{fig:F_f} is the intrinsic SED with the effective temperature set to 8995 K \citep{1994ApJ...424..237S} and the surface gravity set to log$(g) = 4.2$ based on the spectral type\footnote{V410 Anon9 is an A2 type main-sequence star, of which the surface gravity log$(g)$ is supposed to be under 4.2878 (the typical value for an A5 type main-sequence star, \citealt{2000Book}) and over 4.1378 (A0, \citealt{2000Book}). 
We choose the value of 4.2 to represent the surface gravity of V410 Anon9. }.

\citet{2011A&A...526A.103D} indicated a solar metallicity for the nearest large star-forming region Taurus-Auriga association where the tracer V410 Anon9 is located.
As a result, we adopt the protosolar abundances \citep{2009ARA&A..47..481A} and assume a typical value of $f_{\rm cs} = 0.3$ for the mass ratio of graphite to silicate, which means that all the Si are constrained in silicate dust and the fraction of gas-phase carbon is 50\%.
With 56 values of $\alpha$ ($0.5 \le \alpha \le 6.0$ with 0.1 dex steps) and 30 values of $a_c$ ($0.01 \le a_c \le 0.30$ with 0.01 dex steps) considered, a grid of model extinction curves with the same KMH dust distribution for silicate and graphite is constructed for V410 Anon9 after several pre-experiments\footnote{
It is found that similar results could be derived from the dust models with the same dust size distribution and the independent dust size distribution.}
for reducing the number of the parameters and simplifying the calculation.

With the combination of the intrinsic SEDs and model extinction curves, the model SEDs for V410 Anon9 can be written as $F^{\rm mod}_{\lambda} (\alpha_{\rm sil},~a_{c\rm sil},~{\rm log}~T_{\rm eff},~{\rm log}~g)$ [or $F^{\rm mod}_{\lambda} (\alpha_{\rm gra},~a_{c\rm gra},~{\rm log}~T_{\rm eff},~{\rm log}~g)$ for the same dust size distribution].


\subsection{The extinction tracer in the diffuse region}

To explore how the improved pair method is applied to different interstellar environments, we choose a typical extinction tracer in the diffuse region for comparison.  
Located in the stellar association Cygnus OB2, Cygnus OB2  {\#}12 is a Galactic B3-5 hypergiant ~\citep{2013ARep...57..527C,2012A&A...541A.145C}. 
The extinction law toward Cyg OB2 {\#}12 described by standard grain models derived from the former study \citep{2015MNRAS.449..741W} shows consistency to the mean diffuse-ISM curve.
Because of its fairly high brightness ($1.9 \times 10^{19} L_{\odot}$, \citealt{2012A&A...541A.145C}), it can be used as a popular extinction tracer in the diffuse region. 
Many researches on extinction law toward Cyg OB2 {\#}12 have been carried out and indicate that its reddening is higher ($A_V \approx 10.1 $ mag, \citealt{1957PASP...69..239S,2007ApJ...664.1102K,2015MNRAS.449..741W,2020ApJ...895...38H}) than the average reddening of bright stars in the association \citep{2003ApJ...597..957H,2007ApJ...664.1102K,2010ApJ...713..871W,2012ApJS..202...19G,2015MNRAS.449..741W}.
Although recent studies suggest that Cyg OB2 \#12 may be the member of a binary \citep{2014AJ....147...40C,2016MNRAS.458..491M,2017ApJ...845...39O} and the system may even consist of a third member \citep{2016MNRAS.458..491M}, we do not deeply explore its properties as the member in a binary star system in this work and ignore the influence of its possible companion(s).

\subsubsection{The observed SED}\label{subsubsec:cyg_obs}

The observed SEDs for Cyg OB2 \#12 are presented in the lower panel of Figure \ref{fig:sed}.
The spectrum is from the LAMOST \citep{2012RAA....12..723Z, 2012RAA....12.1197C}. The photometric data used in this work are from GAIA DR2 \citep{2018yCat.1345....0G} in $G_{BP}, G, G_{RP}$ bands, the 2MASS \citep{2003yCat.2246....0C} in $J, H, K_S$ bands, the XMM-Newton Serendipitous Ultraviolet Source Survey 4.1 (XMM-SUSS 4.1, \citealt{2019yCat.2356....0P}) in $UVW1$ bands, the MSX6C Infrared Point Source Catalog \citep{2003yCat.5114....0E} in $D, C, A, B_2$ bands, the InfraRed Astronomical Satellite Point Source catalog/Faint Source catalog (IRAS PSC/FSC, \citealt{2015AandC....10...99A}) in $IRAS12, IRAS25$ bands, the AKARI \citep{2010AandA...514A...1I} in $S9W, L18W$ bands, and in $U$ \citep{2015MNRAS.449..741W}, $B$ \citep{2016MNRAS.458..491M}, $V$ \citep{2016MNRAS.458..491M}, and $VT$ \citep{2008PASP..120.1128O} bands.

The observed photometric points for Cyg OB2 \#12 are listed in Table \ref{tab:photometry} and are applied to calculate the extinction law.


\subsubsection{The model SEDs}\label{subsubsec:cyg_mod}

For Cyg OB2 \#12, we also construct the model intrinsic SEDs with 47 values of the effective temperature ($4000 \le T_{\rm eff} \le 50000$ K with 1000 K steps), 17 values of surface gravity ($1.00 \le {\rm log}(g) \le 5.00$ dex with 0.25 dex steps) and the solar abundances as mentioned in Section \ref{subsubsec:v410_mod}.
The intrinsic SEDs from the ATLAS9 stellar model atmosphere are plotted in the lower panel of Figure \ref{fig:F_f} with gray solid lines.
The black dotted line in the lower panel of Figure \ref{fig:F_f} is the intrinsic SED with the stellar parameters [$T_{\rm eff} = 13700^{+800}_{-500}$ K, log$(g) = 1.70^{+0.08}_{-0.15}$] estimated by \citet{2012A&A...541A.145C}.

\citet{2015ApJ...811..110W} showed that the extinction in the line of sight over the accessible wavelength range is well described by standard grain models that fit the mean diffuse-ISM curve.
The element abundances of the diffuse region in the MW are consequently adopted: Si/H $\approx 32$ ppm, C/H $\approx 300$ ppm, and it is assumed that the mean molecular weight of silicate is 170 per Si \citep{1994ApJ...422..164K}.
We derive that the mass ratio of graphite to silicate is $f_{\rm cs} =m_{\rm gra}/m_{\rm sil} = (300 \times 12)/(32 \times 170) \approx 2/3$.
$a_{c\rm sil}$ and $a_{c\rm gra}$ in the KMH model are fixed to 0.25 $\mu {\rm m}$ as mentioned in \citet{2022ApJS..259...12W}, which is the maximum cutoff value of dust size in the MRN model \citep{1977ApJ...217..425M}. 
A grid of model extinction curves for Cyg OB2 \#12 is thus constructed with 56 values of $\alpha_{\rm sil}$ ($0.5 \le \alpha_{\rm sil} \le 6.0$ with 0.1 dex steps) and 56 values of $\alpha_{\rm gra}$ ($0.5 \le \alpha_{\rm gra} \le 6.0$ with 0.1 dex steps) after the pre-experiments\footnote{We also adopted four free parameters ($\alpha_{\rm sil},~a_{c{\rm sil}},~\alpha_{\rm gra},~a_{c{\rm gra}}$) for the fitting, and derived the similar results as those derived with two free parameters ($\alpha_{\rm sil},~\alpha_{\rm gra}$).
However, by comparing the values of the Akaike information criterion for the model with four parameters and that with two parameters, we would prefer the latter model in this work for the better goodness of fitting.}.

With the intrinsic SED extinguished by the model extinction curves, a grid of model SEDs [$F^{\rm mod}_{\lambda}(\alpha_{\rm sil},\alpha_{\rm gra})$] for Cyg OB2 \#12 can be constructed.

\section {Results and discussions} \label{sec:re}

\subsection{Extinction laws toward V410 Anon9 and Cyg OB2 \#12} \label{subsec:ext_law}

The extinction results for V410 Anon9 and Cyg OB2 \#12 are listed in Table \ref{tab:re}.
The fitting parameters and the 1$\sigma$ uncertainties are tabulated based on the 50\%, 16\%, and 84\% values of the marginalized 1D posterior probability distribution functions generated from the EMCEE results.
The fifth column shows the residual sum of squares (RSS) for both tracers calculated from the derived parameters.
The sixth to eighth columns in Table \ref{tab:re} compare the corresponding $R_V$, $E(B-V)$ and $A_V$ derived in this work and those in some former studies. 
$R_V$ in this work is derived from the extinction curve 
\begin{equation}
R_V = A_V/E(B-V) = 1/(A_B/A_V-1),
\end{equation}
$E(B-V)$ is calculated by 
\begin{equation}
E(B-V) = 2.5{\rm log}[(F^{\rm int}_BF^{\rm mod}_V)/(F^{\rm int}_VF^{\rm mod}_B)],  
\end{equation}
and then 
\begin{equation}
A_V = R_V \times E(B-V).  
\end{equation}
The Levenberg-Marquardt method is also applied to fit the models to the observed data for comparison, and the same fitting parameters as those derived from the EMCEE fitting code can be derived.
The best-fitting model SEDs for V410 Anon9 and Cyg OB2 \#12 are compared with the observed photometry in Figure \ref{fig:F_f}.

The $A_V$ value derived for V410 Anon9 is about 1.9\% smaller than that in \citet{1994ApJ...424..237S}, which was calculated from the observed magnitude and the intrinsic magnitude in the V band inferred based on the spectral type.
The minor deviation may be due to the different effective wavelengths in the $V$ band adopted in this work and \citet{1994ApJ...424..237S}.
While the derived $A_V$ value for Cyg OB2 \#12 is similar to those in the former studies (e.g., \citealt{1991MNRAS.249....1T,2016MNRAS.458..491M}), and the tiny differences are attributed to different observed and intrinsic data adopted.
The consistency of the extinction results derived in this work and the former studies shows the practicability and reliability of the improved pair method proposed in this work for calculating the extinction law toward individual sight line.

\begin{figure}
      \centering
     \includegraphics[width=10cm, angle=0]{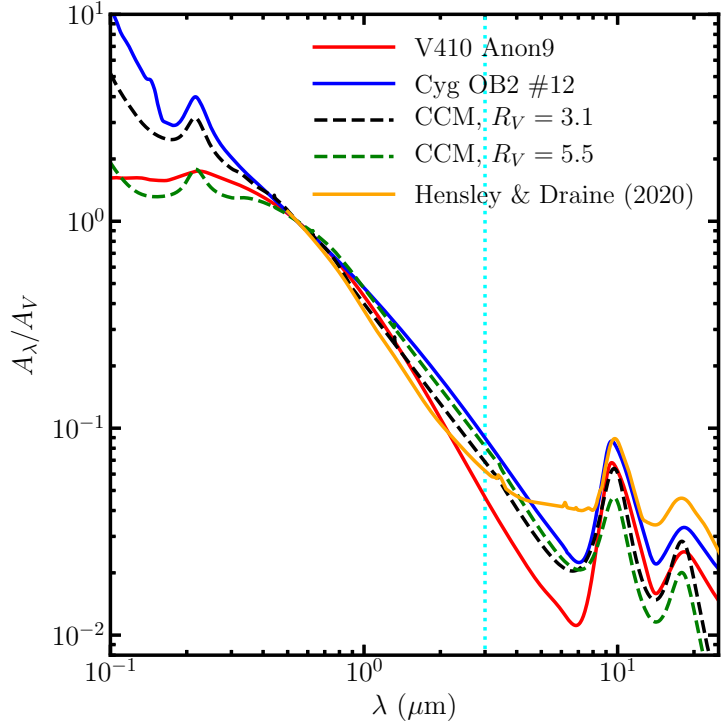}
      \caption{The extinction curves toward V410 Anon9 (red solid line) and Cyg OB2 \# 12 (blue solid line) derived in this work compared with the CCM model.
      The orange line shows the extinction curve toward Cyg OB2 \#12 from \citet{2020ApJ...895...38H}.
      The cyan dotted line indicates $\lambda = 3~\mu{\rm m}$.
      The extinction curves derived in this work at $\lambda < 3~\mu{\rm m}$ are reliable and effective (see Section \ref{subsec:ext_law} for the details).}\label{fig:com_ext}
      \end{figure}

We present the derived extinction curves in Figure \ref{fig:com_ext}.
The comparison of the two extinction curves will be discussed in the next subsection.
Although all dust models for the diffuse ISM predict an extinction curve steeply declines with $\lambda$ at $1~\mu{\rm m} < \lambda < 7~\mu{\rm m}$ and increases at $\lambda > 7~\mu {\rm m}$ because of the 9.7 $\mu {\rm m}$ silicate absorption feature \citep{1977ApJ...217..425M,1994ApJ...422..164K,2001ApJ...548..296W}, many recent observations suggest that the extinction law in the mid-IR band ($3~\mu{\rm m} < \lambda < 8~\mu{\rm m}$) appears to be almost universally flat or gray in various interstellar environments \citep{1996A&A...315L.269L,1999ESASP.427..623L,2005ApJ...619..931I,2007ApJ...663.1069F,2009ApJ...707...89G,2009ApJ...696.1407N,2011ApJ...737...73F,2013ApJ...773...30W,2015ApJ...811...38W}.
\citet{2015ApJ...811...38W} modeled the extinction curve with $\mu{\rm m}$-size grains together with a mixture of silicate and graphite grains of sizes ranging from a few angstroms to a few submicronmeter and found that the flat mid-IR extinction curve at $\sim 3-8~\mu{\rm m}$ provides a sensitive constraint on very large, $\mu{\rm m}$-sized grains.
The $\mu{\rm m}$-sized grains bring gray extinction at optical wavelengths and the extinction curve from the far-UV to near-IR bands is not affected by the existence of the $\mu{\rm m}$-sized grains.
Although adding the $\mu{\rm m}$-sized grains to the silicate-graphite dust model contributes to a better fit in the mid-IR bands, the model will become complex.
In addition, the mid-IR observations cannot be obtained for all the tracers.
As a result, the silicate-graphite dust model adopted in this work is more efficient and more universal to derive the extinction law from UV to near-IR.
The extinction curves derived in this work presented in Figure \ref{fig:com_ext} are reliable at $\lambda < 3~\mu{\rm m}$, while those at $\lambda > 3~\mu{\rm m}$ may suffer large uncertainty.

The extinction curve toward Cyg OB2 \#12 in this work is compared with that derived in \citet{2020ApJ...895...38H}, showing consistency in the optical bands ($\sim 0.3-0.6~\mu{\rm m}$).
The extinction curve in \citet{2020ApJ...895...38H} from the UV to near-IR bands is extracted from \citet{2019ApJ...886..108F} and the minor difference between the extinction curve derived in this work and that from \citet{2020ApJ...895...38H} in the optical to near-IR bands is within the deviation of mean extinction curve given in \citet{2019ApJ...886..108F}.

\begin{table}
  \bc
  \caption[]{Derived fitting parameters and the corresponding $R_V$, $E(B-V)$, $A_V$ for V410 Anon9 and Cyg OB2 \#12 compared to the previous works. \label{tab:re}}
  \setlength{\tabcolsep}{2.5pt}
  \small
   \begin{tabular}{ccccccccccccc}
    \hline\noalign{\smallskip}
      & $\alpha_{\rm sil}$ & $a_{c\rm sil}$ & $\alpha_{\rm gra}$ & $a_{c\rm gra}$ & $RSS^{a}$ & $R_V$ & $E(B-V)$ & $A_V$ & ref\\
      & & ($\mu{\rm m}$) & & ($\mu{\rm m}$) & & & (mag) & (mag) &\\
    \hline\noalign{\smallskip}
		V410 Anon9 & $0.80^{+0.03}_{-0.03}$ & $0.050^{+0.003}_{-0.003}$ & $0.80^{+0.03}_{-0.03}$ & $0.050^{+0.003}_{-0.003}$ & 1.05 & 4.26 & 2.40 & 10.24 & EMCEE fitting code\\
     & 0.80 & 0.050 & 0.80 & 0.050 & 1.05 & 4.26 & 2.40 & 10.24 &  Levenberg-Marquardt method \\
		 & - & - & - & - & - & -  & & 10.45 & \citet{1994ApJ...424..237S}\\
		Cyg OB2 \#12$^b$ & $\bm{4.65^{+0.03}_{-0.03}}$ & 0.25 & $\bm{3.60^{+0.03}_{-0.03}}$ & 0.25 & 1.11 & 3.00 & 3.36 & 10.08 & EMCEE fitting code\\
		 & 4.70 & 0.25 & 3.60 & 0.25 & 1.11 & 3.00 & 3.36 & 10.08 & Levenberg-Marquardt method \\    
		 & - & - & - & - & - & 3.00 & 3.34 & 10.02 & \citet{2016MNRAS.458..491M} \\
		 & - & - & - & - & - & 3.04 & 3.36 & 10.20 & \citet{1991MNRAS.249....1T} \\          
    \noalign{\smallskip}\hline
  \end{tabular}
  \ec
  \tablecomments{0.86\textwidth}{$^a$ The residual sum of squares $ = \sum [{\rm log}(f_{\rm model})-{\rm log}(f_{\rm observed})]^2$.\\
  $^b$ The same dust size distribution of silicate and graphite is adopted for V410 Anon9.} 
  \end{table}

\subsection{Influence of interstellar environment on the extinction law} \label{subsec:ext_com}

It is generally believed that the interstellar environment has a great influence on the extinction law.
The size of grain in the dense region is larger, while smaller grain distributes in the diffuse region.
V410 Anon9 is a typical extinction tracer in the dense region, and Cyg OB2 \#12 is located in the diffuse region.
The extinction curves derived for both tracers with the typical CCM extinction curves are shown in Figure \ref{fig:com_ext}, from which we can notice a remarkable distinction between the extinction laws toward both tracers, indicating different dust properties.
By comparing the results of both tracers, we can explore how different interstellar environments impact on the properties of extinction.
In addition, the application of the improved pair method proposed in this work to different interstellar environments can also be proved.

The extinction curve derived for V410 Anon9 is relatively flatter and shows a weaker 2175 {\rm \AA} bump feature indicating large dust size, which is in line with the properties of dust in the dense interstellar environment ($R_V \approx 5.5$).
For Cyg OB2 \#12, the extinction curve exhibits consistency to the mean extinction curve of the MW ($R_V \approx 3.1$), and the 2175 \,{\rm \AA} bump in the extinction curve is obvious.
The significantly different absorption features at $\sim$ 2175 \,{\rm \AA} in both extinction curves indicate that the dust size toward V410 Anon9 is larger than that toward Cyg OB2 \#12.

\begin{figure}
  \centering
 \includegraphics[width=10cm, angle=0]{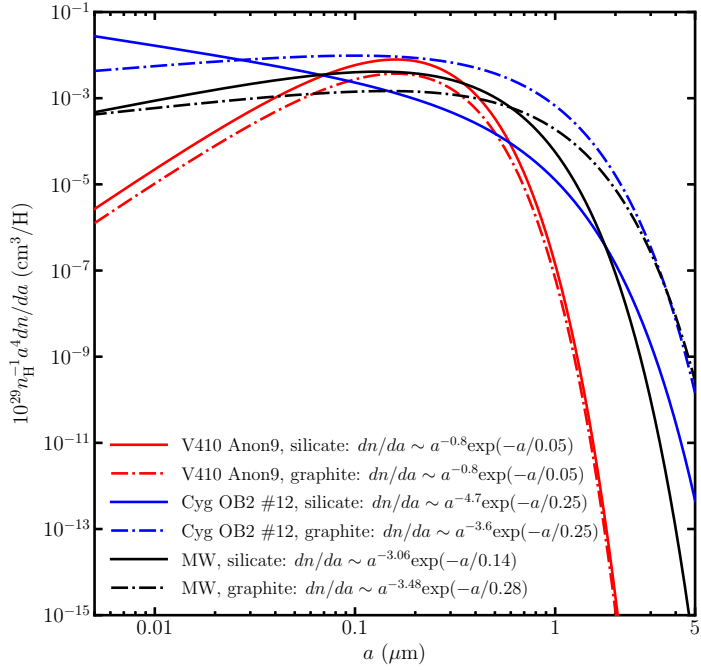}
  \caption{The dust size distributions toward V410 Anon9 and Cyg OB2 \# 12 derived in this work as well as the KMH dust size distributions of the silicate and graphite in the MW.\label{fig:com_dis}}
  \end{figure}

In addition to the reliable extinction results, the parameters included in the dust model obtained from the improved pair method can also be used to analyze the dust properties quantitatively, which cannot be achieved by the mathematical extinction laws.
We present the dust size distributions toward V410 Anon9 and Cyg OB2 \#12 in Figure \ref{fig:com_dis} as well as the KMH dust size distributions for silicate and graphite in the MW \citep{1994ApJ...422..164K}.
The equation (8) in \citet{2016P&SS..133...36N} is adopted to calculate the average dust size $\overline{a}$, and we derive that $\overline{a_{\rm sil}} = \overline{a_{\rm gra}} \approx 0.03~\mu\rm m$ for V410 Anon9 and $\overline{a_{\rm sil}} \approx 0.006~\mu \rm m$, $\overline{a_{\rm gra}} \approx 0.007~\mu\rm m$ for Cyg OB2 \#12.
Compared with $\overline{a_{\rm sil}} \approx 0.008~\mu \rm m$ and $\overline{a_{\rm gra}} \approx 0.007~\mu\rm m$ for the MW with the KMH distributions calculated from the same method, the average dust size of Cyg OB2 \#12 is similar to that of the MW and is smaller than that of V410 Anon 9.

\subsection{Influence of stellar parameters on the extinction law} \label{subsec:st_ext}

The stellar parameters derived with the improved pair method for V410 Anon9 is that ${\rm log}(T_{\rm eff}) = 3.92^{+0.004}_{-0.005}$ and ${\rm log}(g) = 4.24^{+0.172}_{-0.169}$.
The derived value of ${\rm log}(g)$ is reasonable for the A-type main sequence star \citep{2000Book}, while the derived ${\rm log}(T_{\rm eff})$ is lower than that [${\rm log}(T_{\rm eff}) \approx 3.95$] in \citet{1994ApJ...424..237S}.
The intrinsic SED with the derived stellar parameters is presented in the upper panel of Figure \ref{fig:F_f} with a blue solid line compared to that with log($T_{\rm eff}$) = 3.95 and log$(g)$ = 4.2.
For Cyg OB2 \#12, we derived that ${\rm log}(T_{\rm eff}) = 4.14^{+0.056}_{-0.030}$ and ${\rm log} (g) = 1.60^{+0.125}_{-0.032}$, which corresponds to the estimated stellar parameters [${\rm log}(T_{\rm eff}) = 4.14$ and ${\rm log} (g) = 1.7$] in \citet{2012A&A...541A.145C} and proves that the method proposed in this work can also be used to infer the stellar parameters.

\begin{figure}
  \centering
 \includegraphics[width=15 cm, angle=0]{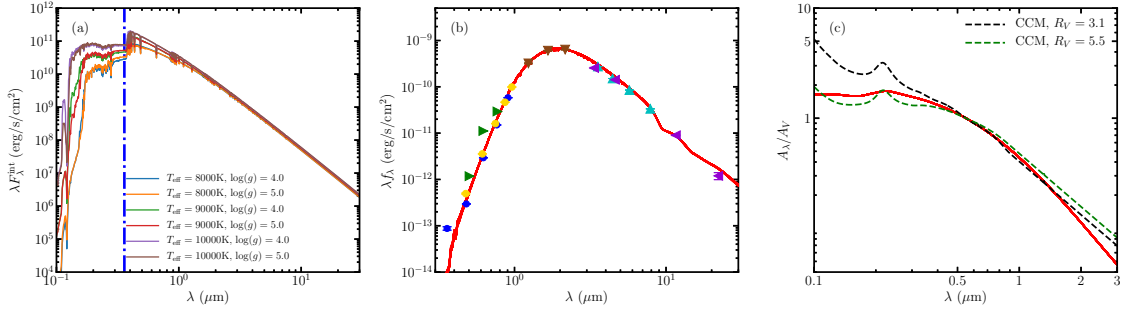}
  \caption{Test for the sensitivity of the results to the stellar parameters. 
  Panel (a): intrinsic spectra derived from the ATLAS9 stellar model atmosphere with different stellar parameters.
	The blue dash line indicates the shortest effective wavelength of the observed photometry for V410 Anon9 ($\lambda = u_{\rm eff}$) in this work.
  Panel (b): best-fitting model SEDs derived from different possible stellar parameters adopted for V410 Anon9 compared with the observed photometry.
  Panel (c): extinction curves of V410 Anon9 derived with different possible stellar parameters. \label{fig:diff_st}}
  \end{figure}

As we know, different stellar parameters lead to different intrinsic SEDs, thus may lead to different derived extinction curves.
As a result, it is necessary to find out the sensitivity of the extinction law to the stellar parameters.
To show the influence of the stellar parameters on the intrinsic SEDs, we plot the intrinsic SEDs derived from the ATLAS9 stellar model atmosphere with different stellar parameters in Figure \ref{fig:diff_st} (a).
The blue dashed-and-dotted line indicates the lower limit of effective wavelengths of the observed photometry that we adopt for V410 Anon9 in this work ($u$ band).
From Figure \ref{fig:diff_st} (a), we can find that log($g$) has little effect on the intrinsic flux, while the intrinsic flux increases with the increase of $T_{\rm eff}$, especially in UV and shorter bands.

As the spectral type of tracer V410 Anon9 is A2, we adopt some other possible values of log($g$) and $T_{\rm eff}$ for the A-type main sequence star [$7300{\rm K} < T_{\rm eff} < 10000~{\rm K}$, $4.14 < {\rm log}(g) < 4.34$ according to \citet{2000Book}] to test the sensitivity of the extinction law to the stellar parameters for V410 Anon9.
 
\begin{figure}
      \centering
     \includegraphics[width=10 cm, angle=0]{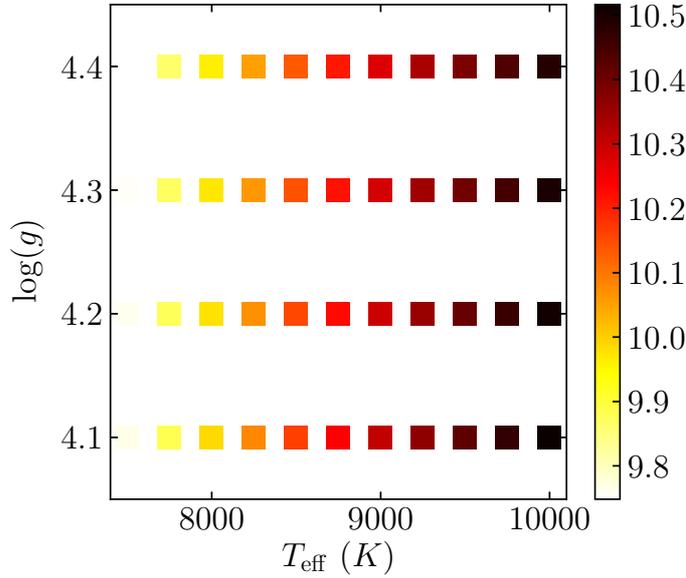}
      \caption{The values of $A_V$ derived for A-type main sequence star with different stellar parameters.\label{fig:diff_av_dis}}
      \end{figure}

We list the results derived with the EMCEE fitting code for each pair of the stellar parameters in Table \ref{tab:diff_st}, which are the same as those derived with the Levenberg-Marquardt method, and compare the model SEDs with the observed data in Figure \ref{fig:diff_st} (b).
The extinction curves derived with different stellar parameters are presented in Figure \ref{fig:diff_st} (c), while the derived values of $A_V$ ranging from 9.75 mag to 10.52 mag\footnote{
The color changes mimic the effect of extinction, so there are degeneracies between the effective temperature and the extinction values. 
However, thanks to the wide wavelength coverage of the observations, the results derived by the method proposed in this work are not affected by the degeneracy between the effective temperature and the extinction values according to the parameter sensitivity tests in \citet{2022ApJS..259...12W}.} are plotted in Figure \ref{fig:diff_av_dis}.
It is illustrated that the small variation of the stellar parameters in this work has little effect on the extinction curves and dust properties for V410 Anon9, because the intrinsic spectra at $\lambda > \lambda_u$ are not strongly influenced by the stellar parameters as presented in Figure \ref{fig:diff_st}.
It is also suggested that even if the stellar parameters could not be obtained directly, we can still derive the extinction law and analyze the properties of the dust based on the spectral type of the tracer.

  \begin{table}
    \bc
    \begin{minipage}[]{150mm}
    \caption[]{The derived fitting parameters ($\alpha$, $a_{c}$) and the corresponding $A_V$ for V410 Anon9 with different possible stellar parameters adopted$^a$.\label{tab:diff_st}}
    \end{minipage}
    \setlength{\tabcolsep}{2.5pt}
    \small
     \begin{tabular}{ccccccccccccc}
      \hline\noalign{\smallskip}
        $T_{\rm eff}$ & log$(g)$ & $\alpha$ & $a_{c}$ & $A_V$\\
        (K) & & & $(\mu{\rm m})$ &(mag)\\            
      \hline\noalign{\smallskip}
      8995 & 4.2 & $0.8_{-0.03}^{+0.03}$ & $0.05_{-0.003}^{+0.003}$ & 10.29 \\
      \hline
      7500 & $4.1$ & $0.8_{-0.03}^{+0.03}$ & $0.05_{-0.003}^{+0.003}$ & 9.77 \\
       & $4.2$ & $0.8_{-0.03}^{+0.03}$ & $0.05_{-0.003}^{+0.003}$ & 9.76 \\
       & $4.3$ & $0.8_{-0.03}^{+0.03}$ & $0.05_{-0.003}^{+0.003}$ & 9.75 \\
       & $4.4$ & $0.8_{-0.03}^{+0.03}$ & $0.05_{-0.003}^{+0.003}$ & 9.75 \\
      8000 & $4.1$ & $0.8_{-0.03}^{+0.03}$ & $0.05_{-0.003}^{+0.003}$ & 9.99 \\
       & $4.2$ & $0.8_{-0.03}^{+0.03}$ & $0.05_{-0.003}^{+0.003}$ & 9.98 \\
       & $4.3$ & $0.8_{-0.03}^{+0.03}$ & $0.05_{-0.003}^{+0.003}$ & 9.97 \\
       & $4.4$ & $0.8_{-0.03}^{+0.03}$ & $0.05_{-0.003}^{+0.003}$ & 9.96 \\
      8500 & $4.1$ & $0.8_{-0.03}^{+0.03}$ & $0.05_{-0.003}^{+0.003}$ & 10.16 \\
       & $4.2$ & $0.8_{-0.03}^{+0.03}$ & $0.05_{-0.003}^{+0.003}$ & 10.15 \\
       & $4.3$ & $0.8_{-0.03}^{+0.03}$ & $0.05_{-0.003}^{+0.004}$ & 10.14 \\
       & $4.4$ & $0.8_{-0.03}^{+0.03}$ & $0.05_{-0.003}^{+0.003}$ & 10.13 \\
      9000 & $4.1$ & $0.8_{-0.03}^{+0.03}$ & $0.05_{-0.003}^{+0.003}$ & 10.3 \\
       & $4.2$ & $0.8_{-0.03}^{+0.03}$ & $0.05_{-0.003}^{+0.003}$ & 10.29 \\
       & $4.3$ & $0.8_{-0.03}^{+0.03}$ & $0.05_{-0.003}^{+0.003}$ & 10.28 \\
       & $4.4$ & $0.8_{-0.03}^{+0.03}$ & $0.05_{-0.003}^{+0.003}$ & 10.27 \\
      9500 & $4.1$ & $0.8_{-0.03}^{+0.03}$ & $0.05_{-0.004}^{+0.003}$ & 10.42 \\
       & $4.2$ & $0.8_{-0.03}^{+0.03}$ & $0.05_{-0.003}^{+0.003}$ & 10.41 \\
       & $4.3$ & $0.8_{-0.03}^{+0.03}$ & $0.05_{-0.003}^{+0.003}$ & 10.4 \\
       & $4.4$ & $0.8_{-0.03}^{+0.04}$ & $0.05_{-0.003}^{+0.003}$ & 10.39 \\
      10000 & $4.1$ & $0.8_{-0.03}^{+0.03}$ & $0.05_{-0.003}^{+0.003}$ & 10.52 \\
       & $4.2$ & $0.8_{-0.03}^{+0.03}$ & $0.05_{-0.003}^{+0.003}$ & 10.51 \\
       & $4.3$ & $0.8_{-0.03}^{+0.03}$ & $0.05_{-0.003}^{+0.003}$ & 10.5 \\
       & $4.4$ & $0.8_{-0.03}^{+0.04}$ & $0.05_{-0.003}^{+0.003}$ & 10.49\\
      \noalign{\smallskip}\hline
    \end{tabular}
    \ec
    \tablecomments{1\textwidth}{$^a$ The first line is the results for V410 Anon9 with $T_{\rm eff}$ from \citet{1994ApJ...424..237S} and log$(g)$ based on the spectral type.}
    \end{table}  

\section{Summary of the improved method} \label{sec:summary}

An improved pair method is proposed in this work based on the traditional pair method.
The improved method combines the intrinsic SEDs from the ATLAS9 stellar model atmosphere and the model extinction curves from the silicate-graphite dust model.
Adopting a stellar model atmosphere to obtain intrinsic spectra is to eliminate the impact of the limited standard stars \citep{2005AJ....130.1127F}.
While extinction curves from the dust model can be applied to various interstellar environments and help to analyze the dust properties.
The pre-construction of the template library for the model SEDs greatly reduces the time required for the calculation.
V410 Anon9 located in the dense region of the MW and Cyg OB2 \#12 located in the diffuse region of the MW are chosen as the extinction tracers, and the derived results are consistent with their interstellar environments and the previous studies.
It is proved that the improved pair method proposed in this work is reliable and efficient to calculate the dust extinction law from UV to near-IR bands toward various interstellar environments.
In addition to the reliable extinction results, the dust properties can be analyzed and the stellar parameters can be inferred with the improved pair method at the same time.
Furthermore, the extinction law beyond the wavelengths of observed data can also be predicted based on the dust model.

However, as mentioned in Section \ref{subsec:ext_law}, the improved pair method may not be applied in the mid-IR band, because the extinction curves derived from the silicate-graphite model in the mid-IR band ($\sim 3-8~\mu{\rm m}$) are not as flat as the extinction law calculated from the recent observations.
Although the $\mu{\rm m}$-sized grain can contribute to modeling the flat mid-IR extinction curve \citep{2015ApJ...811...38W}, it will increase the complexity of the model and thus reduce the universality of the improved method.
As a result, the improved pair method proposed in this work can be used to derive the extinction curve from UV to near-IR bands at present.
In addition, the reliability of the improved pair method depends on the wavelength coverage of the observed data.
Lacking UV or near-IR data may lead to unreliable results \citep{2022ApJS..259...12W}.
To obtain more appropriate extinction curves with the improved pair method, observed data that covers the wavelength from UV to near-IR are needed.

The improved pair method proposed in this work can be applied to large-scale samples in the MW and external galaxies.
Despite the unknown stellar parameters, dozens of extinction curves toward individual sight lines have been derived in M31 \citep{2022ApJS..259...12W}, M33 \citep{2022ApJS..260...41W}, and other external galaxies \citep{2022ApJS..260...41W} based on the spectral types of the extinction tracers. 
In addition, the coming 2 m-aperture Survey Space Telescope (also known as the China Space Station Telescope, CSST, \citealt{2021CSB...111.111C}) will image the sky and will provide us with abundant data in the UV to IR bands with high resolution.
We can combine the improved pair method with the multi-band photometry from the CSST to get a more comprehensive understanding of the dust extinction law and dust properties in the MW and external galaxies.


\normalem
\begin{acknowledgements}
We would like to thank the anonymous reviewer for the helpful and insightful report.
It is a pleasure to thank Prof. Biwei Jiang and Haibo Yuan for the very helpful discussions, and thank Mr. Weijia Gao and Ruining Zhao for the dust model.
We especially acknowledge and thank Prof. Jingyao Hu for the idea about V410 Anon9.
This work is supported by the Natural Science Foundation of China through projects NSFC 12133002, U2031209 and 12203025, Shandong Provincial Natural Science Foundation through project ZR2022QA064, and the CSST Milky Way and Nearby Galaxies Survey on Dust and Extinction Project CMS-CSST-2021-A09.
\end{acknowledgements}
  
\bibliographystyle{raa}
\bibliography{bibtex}{}

\end{document}